\documentclass[%
 reprint,
 showkeys,
 showpacs,
 amsmath,amssymb,
 aps,
prb,
]{revtex4-1}

\usepackage{graphicx}
\usepackage{dcolumn}
\usepackage{bm}

\begin{document}

\preprint{APS/123-QED}

\title{Atomic structure and mechanical properties of carbyne\\}

\author{A. Timoshevskii}
 \email{tim@imag.kiev.ua}
\author{S. Kotrechko}
\author{Yu. Matviychuk}
\affiliation{%
 G.V. Kurdyumov Institute for Metal Physics of the National Academy of Sciences of Ukraine, 03142 Kiev, Ukraine\\
}%

\date{\today}

\begin{abstract}
The atomic structure and mechanical properties of the carbyne (monatomic linear chains), containing from 2 to 21 carbon atoms, are theoretically investigated by  \textit{ab-initio} methods. We demonstrate the existence of a stable cumulene-structure in the inner part of chains with the number of atoms $N\geqslant10$. We present a general stress-strain diagram of chains until the moment when they break, which enables to determine their strength, elasticity and fragility. For chains with $N\geqslant 4$, the relationship between the strength of the chain and the binding energy of the edge atom in the chain is established. The existence of scale-effect and "even-odd" effect for such properties as strength, elasticity and fragility is observed. We demonstrate that the 5-atom carbon chains show the maximum strength value.
\end{abstract}


\maketitle


\section{\label{sec:level1}Introduction}

Carbynes (monatomic linear chains of carbon) have recently attracted much attention due to their unusual physical properties  ~\cite {Lou1, Ravagnan2, Cinquanta3, Rinzler4}, and promising applications ~\cite{Durgun5, Wang6, Erdogan7, Rinzler8, Lang9, Yazdani10}. The potential implementations of these unique functional properties essentially depend on the strength and elasticity of the carbine. Moreover, the possibility of obtaining carbine by unraveling out of nanotubes or graphene sheets is governed by it mechanical properties ~\cite{Durgun5, Wang6, Ragab11, Ataca12}. Therefore, a lot of studies concernig the investigation of strength and stability of monatomic carbon chains have appeared recently ~\cite{Kavan13, Casari14, Jin15, Liu16}. The results of \emph{ab-initio} calculations are still forming the main source of information about the mechanical properties of carbyne. Only recently the experimental results of tests to determine the tensile strength of carbon atomic chains using a high-field method have been published ~\cite{Kotrechko17, Mikhailovskij18}, and extremely high level of strength of these chains, which exceeds 270 GPa was found ~\cite{Mikhailovskij18}. Moreover, their high-field-evaporation stability was ascertained.

At the same time, most of the studies do not present the results of targeted investigations of the dependence of the strength of chains on their atomic structure. The mechanical properties data is usually the auxiliary one when analyzing electronic, magnetic and other functional properties. Currently, to the best of our knowledge, this information is strongly contradictory in literature. For example, the data on the strength of monatomic carbon chains in various articles differs by the order of magnitude ~\cite{Huang19}], and in most cases, the properties of chains of \emph{infinite} length are investigated. However, the structure and properties of a finite carbon chain (carbyne) differ significantly from those for an infinite chain ~\cite{Fan20, Cahangirov21}. The present study is focused on \emph{ab-initio} simulations of carbyne atomic structures with a detailed analysis of their strength, elasticity and fragility.

\section{\label{sec:level1}The subject and the methods of investigation}
As it was shown in ~\cite{Fan20, Cahangirov21}, the interatomic distance in the chains of different lengths is not constant and depends on the number of atoms in the chain. In the present study we carry out a detailed analysis of this effect using \emph{ab-initio} calculations of the electronic and atomic structure of chains of different lengths with even and odd number of atoms. The number of atoms in the chain was varied from 2 to 21.

The mechanical properties and total energies of carbon chains were calculated using the quantum-espresso program package ~\cite{Giannozzi22}, which uses planewaves as a basis set. We employed ultrasoft pseudopotentials, which were custom-generated for carbon atoms using the vanderbilt code (v 7.3.4)~\cite{Giannozzi23}. The planewave basis cutoff of $E_{cut} = 450 eV$ was used in all calculations. To check the pseudopotentials and the basis cutoff value, a series of test calculations for infinite chains with the structure of cumulene and polyyne were performed. The obtained interatomic distances and total energies were consistent with the results of the previous study ~\cite{Cahangirov21}. To calculate and visualize the spatial distribution of the electron density in the chains, an all-electron  Linearised Augmented Plane Wave (\emph{LAPW}) method was employed, as implemented in the WIEN2K program package~\cite{Blaha24}. The generalized gradient approximation (GGA) for exchange-correlation functional was used in all calculations ~\cite{Perdew25}.

As the results of \emph{ab-initio} calculations, the  dependence of the tensile force $F$ on the value of total chain strain $e$ was determined, as well as the values of the bond strain $\varepsilon$ between the first and the second atoms from the edge of the chain (Fig. ~\ref{fig:fig1}). The force acting on the edge atom was calculated as:

\begin{figure*}
\includegraphics{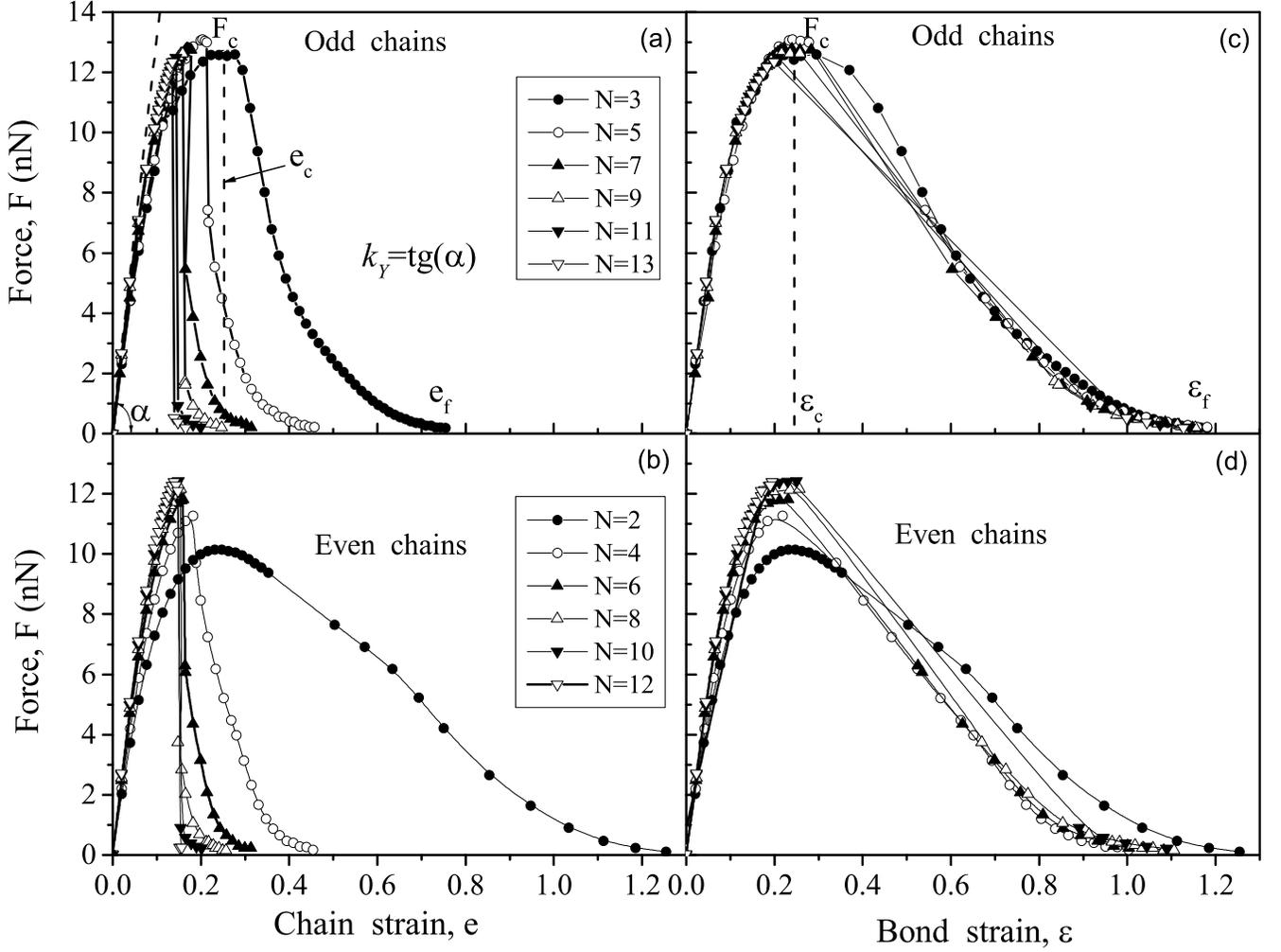}
\caption{\label{fig:fig1} The dependence of force $F$ on strain $e$ for the whole chain with odd (a) and even (b) number of atoms, and on the value of strain $\varepsilon$ at the edge atomic bond in "odd" (c) and "even" (d) chains: $F_c$ is critical stress for instability of atomic bond (bond strength); $\varepsilon_c$ and $e_c$ are critical strain of instability of atomic bond and the chain as a whole, respectively; $\varepsilon_f$ and $e_f$ are fracture strain for atomic bond and chain, respectively; $k_Y$ is coefficient of elasticity of the chain.
}
\end{figure*}

\begin{equation}
 F=\frac{dE}{da},%
\end{equation}

where $E$ is the total energy of the system; $a$ is the current distance between the first and the second atoms (length of the edge bond).

\begin{figure}
\includegraphics{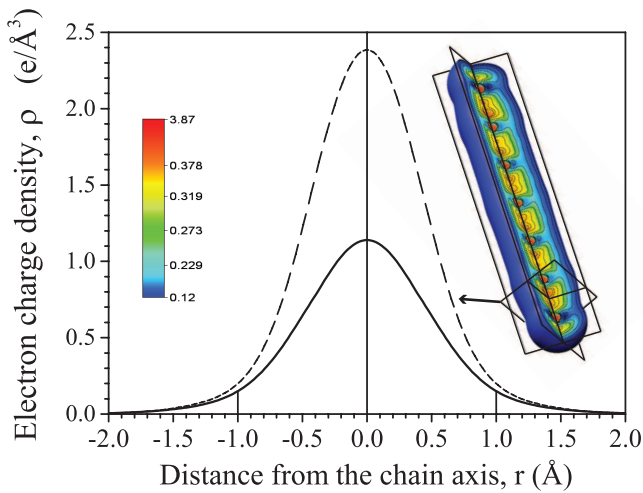}
\caption{\label{fig:fig2} (Color online) The distribution of the electron density in the bulk of the chain and its cross-section passing through the middle of the edge atomic bond in the unloaded state (dashed line) and at the moment of instability of the chain (solid line).}
\end{figure}

The strain of the whole chain, $e$, was estimated as:

\begin{equation}
 e=\ln\left(\frac{l}{l_0}\right),%
\end{equation}

where $l_0$ and $l$ are the initial and current chain lengths. The strain of interatomic bonds between the first and the second atoms was calculated as:

\begin{equation}
 \varepsilon=\ln\left(\frac{a_i}{a_0}\right),%
\end{equation}

where $a_0$ and $a_i$ are the initial and current length of the edge bond, respectively. The force-strain dependencies (Fig. ~\ref{fig:fig1}) enabled us to estimate the chain strength $F_c$, which was calculated as a maximum force value at the moment of chain instability. The values of critical strains of instability of the edge atom bond with the chain $\varepsilon_c$ (Fig. ~\ref{fig:fig1}c, d), as well as the value of critical strain for the whole chain $e_c$ (Fig.~\ref{fig:fig1} a, b) were obtained. The values of the elasticity coefficient (stiffness) $k_Y$ and of the elastic $Y$ modulus were calculated as:

\begin{equation}
 k_Y=\left.\frac{dF}{de}\right|_{e=0},%
\end{equation}

\begin{equation}
 Y=\frac{k_Y}{S},%
\end{equation}

where $S$ is the effective cross-sectional area of the chain, through which the atoms interact with each other. To estimate this value, the calculated distribution of electron density in the chain have been used. Figure ~\ref{fig:fig2} shows the spatial distribution of the electron density for a chain of seven atoms in the unloaded state, calculated by the \emph{LAPW} method. The figure shows the change in the density of the electron charge in the cross-section of the chain in the unloaded state and at the critical strain $e_c$ of chain instability. The cross-section for the unloaded state was chosen in the middle of the distance between the two edge atoms. For the critical state of the chain the cross-section was chosen at a distance from the edge atom, at which the magnitude of the electron charge in a plane perpendicular to the axis of the chain, reaches its minimum value.

This approach enables us to give the lower estimate for the transverse size of the area of the force interaction between the first and the second atoms in the chain, which value was  found to be $d\approx2${\AA}. We note that the same effective diameter of the chain has been used in ~\cite{Mazilova26} to simulate the formation of carbyne by unraveling. This value coincides with the effective magnitude $d$ obtained in ~\cite{Gang27} to analyze the thermal conductivity of chains. Furthermore, this value enables to obtain the calculated value of strength of carbyne, which correlates reasonably well with its experimental value, as established in ~\cite{Mikhailovskij18}.

\section{Results and Discussion}

\begin{figure}
\includegraphics{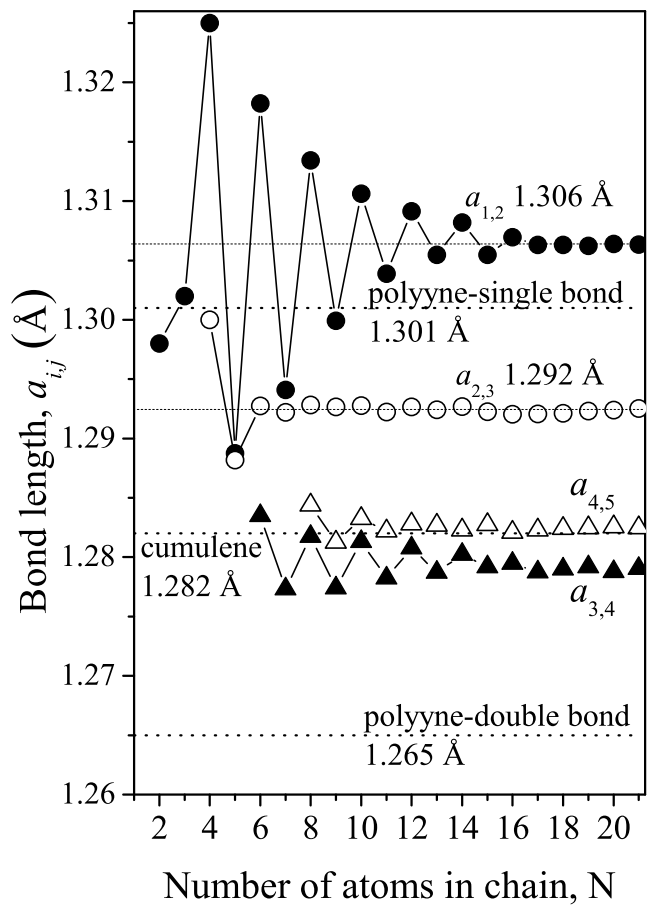}
\caption{\label{fig:fig3} The dependence of the length of interatomic bond, $a_{i,j}$, on the number of atoms in the chain $N$.
}
\end{figure}

Figure ~\ref{fig:fig3} shows the distance between the two nearest neighboring atoms $a_{i,i+1}(i=1,\dots,N)$ in carbon chains of different lengths. According to these results, the distance between the atoms depends on their positions in the chain, as well as on the length of the chain itself. This represents the principle difference between the atomic structure of carbyne and that of infinite carbon chains. The distance between the two outer atoms (the first and the second - $a_{1,2}$) appeared to be the largest one, while the smallest distance was observed between the third and the fourth atoms. In chains with less than 16 atoms, the distance $a_{1,2}$ depends on both the total number of atoms in the chain, and on whether this number is odd or even. Therefore the "scale" and "even-odd" effects occur simultaneously. Our calculations showed that in carbynes with more than 10 atoms, the interatomic distances within the chain (starting from the fourth atom) are equal to those in cumulenes, i.e. the internal structure of carbyne is a cumulene one. It should be noted that the cumulene structure is unstable in the chains of infinite length, and for such chains a polyyne structure is energetically favorable ~\cite{Liu16, Artyukhov28}. Therefore, the existence of a cumulene structure in the inner part of the chain is a specific characteristic of the atomic structure of carbyne. The presence of the edge atoms appears to be the reason for stability of a cumulene-structure in the central part of the finite length chains.

\begin{figure}
\includegraphics{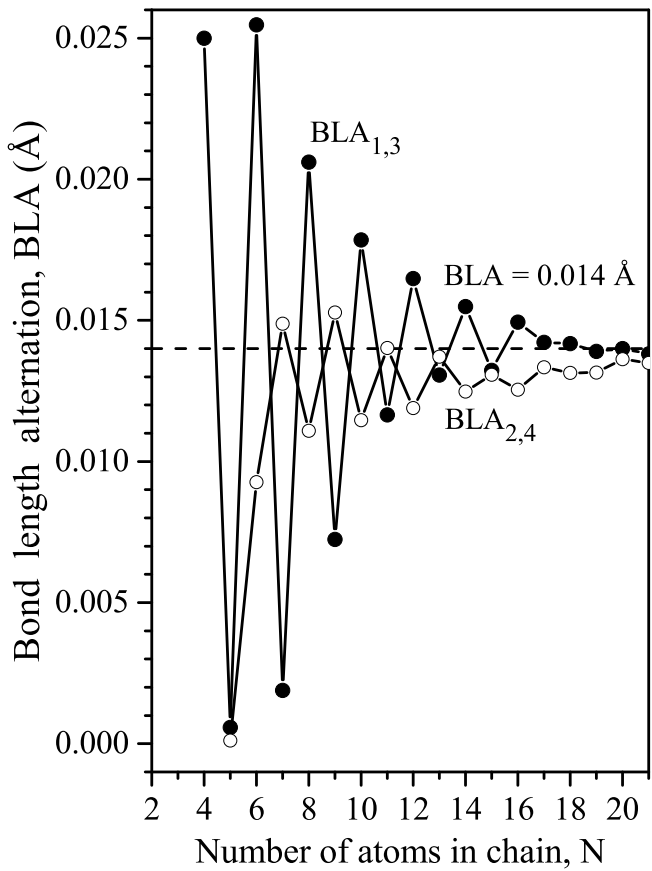}
\caption{\label{fig:fig4} The dependence of the value of bond length alternation  (BLA) on the number of atoms in the chain $N$: $BLA_{1,3}$ and $BLA_{2,4}$ are the values of $BLA$ for atomic bonds $a_{1,2}$ and $a_{2,3}$; $a_{2,3}$ and $a_{3,4}$ respectively.}
\end{figure}

Similar to the case of the infinite chains, the ratio of the bond lengths in finite chains can be characterized using the $BLA_{i,j}$ (bond length alternation) quantity. The regularities of changes in $BLA_{1,3}=a_{1,2}-a_{2,3}$ and $BLA_{2,4}=a_{2,3}-a_{3,4}$ values in chains of different lengths are presented in Figure ~\ref{fig:fig4}. General regularity of changes in the $BLA_{1,3}$ and $BLA_{2,4}$ values lies in the fact that the chains with an even number of atoms, show maximum values of these quantities, while the chains having an odd number of atoms - the minimum ones. Thus, the opposite trends in BLA change with the growth of the number of atoms $N$ are observed. In even-numbered chains the values of $BLA_{1,3}$ and $BLA_{2,4}$ decrease with $N$ (at $N\geq6$), while in odd-numbered chains, they are increasing. At $N\geq19$, these values no longer depend on the total number / parity of atoms and they approach a constant value of $BLA\approx0.014${\AA}. This value is significantly smaller than the BLA for the infinite polyyne chain (0.070-0.090{\AA}) ~\cite{Cahangirov21}.

\begin{figure}
\includegraphics{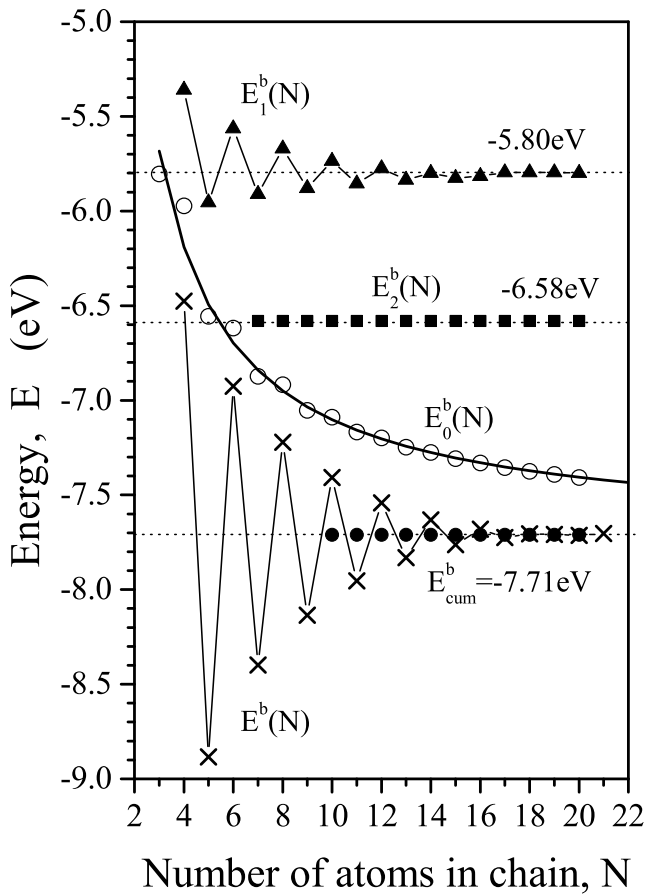}
\caption{\label{fig:fig5} The dependence of the binding energy on the number of atoms: $E_1^b(N)$ and $E_2^b(N)$ - for the first and second atoms from chain edge, $E_{cum}^b$ - for inner atoms, $E^b(N)$ - energy of separation, calculated on the equation (8), $E_0^b(N)$ - the average binding energy per atom (solid line - calculated on the equation (11)).
}
\end{figure}

Significant changes in the interatomic distances along the chain (Fig. ~\ref{fig:fig3}) indicate that the interatomic interaction energy also varies depending on the position of atom in the chain. Therefore, to describe interatomic interaction in the carbyne, the binding energy of each atom with the rest of the chain, $E_i^b(N)$, should be estimated. The total energy of a finite chain, $E(N)$, consisting of $N$ atoms, can be represented as:

\begin{equation}
 E(N)=\sum_{i}^N E_i^b(N)+NE_{at},%
\end{equation}

where $E_{at}$ is the energy of a free carbon atom. In carbyne structures with the number of atoms $N\geq6$,  three types of carbon atoms with different binding energies can be identified. Our analysis of the interatomic distances (Fig.~\ref{fig:fig3}) shows that the first (from the edge) atom should have the lowest binding energy, as the distances $a_{1,2}$ between the first and the second atoms are the largest ones in carbyne. At the same time, the atoms in the central part of the carbyne are expected to have the highest values of the binding energies, as the distances between those atoms are approximately the same and are close to the bond length in cumulenes. The second atoms from edge are expected to have intermediate values of the binding energies. This allows us to represent total energy (6) as follows:

\begin{equation}
 E(N)=NE_{at}+2\left[E_1^b(N)+E_2^b(N)\right] + (N-4)E_{cum}^b,%
\end{equation}

where $E_{at}$ is the energy of a free carbon atom, $E_1^b(N)$ and $E_2^b(N)$ are binding energies of first and second atoms from chain edge, and $E_{cum}^b=-7.71eV$ is a binding energy of the carbon atom in cumulene structure. The total energy calculations of finite chains with different numbers of atoms allowed us to evaluate the binding energy of carbon atoms with different location inside the chain. The results of the calculations are presented in Fig.~\ref{fig:fig5}. In the first approximation, the value of $E_1^b(N)$, in chains with $N\geq16$ is equal to the average binding energy of carbon atom in a chain of three atoms $E_1^b=-5.80 eV$ (Fig.~\ref{fig:fig5}). Such approximation is based on the fact that interatomic distances in the chain are close to the value $a_{1,2}$ in chain with number of atoms $N\geq16$ (Fig.~\ref{fig:fig3}). Binding energy of the near-edge atom ($E_2^b=-6.58 eV$) was obtained from the expression (7) for the total energy of the chain containing 16 atoms using the $E_1^b=-5.80 eV$ value. Interestingly, the average energy of the atom in the 5-atom chain is equal to 6.55eV and is close to the value of $E_2^b$. This can be explained by the fact that the interatomic distances in a 5-atom chain are equal, and are close to the value of $a_{2,3}$ in the chains with $N\geq16$ (see Fig. ~\ref{fig:fig3}). According to our calculations, the value of $E_1^b$, depends both on the total number of atoms in carbyne and on the parity of this number (see Fig. ~\ref{fig:fig5}). In chains with an odd number of atoms, the values of $E_1^b$ are greater in magnitude and decrease with increasing number of atoms in the chain, approaching the value of $E_1^b=-5.80 eV$. The situation is different for even-numbered chains, where binding energy of the edge atoms increase with $N$ (see Fig. ~\ref{fig:fig5}). It is interesting to note the fact that the stronger odd-numbered carbynes are insulators, and even-numbered ones are conducting systems. However, a detailed study of the electronic structure of carbyne systems is beyond the scope of the present paper and should be a topic of a separate investigation.

The data presented in Fig. ~\ref{fig:fig5} demonstrates that the atomic bonds of the edge atoms are the weakest ones. As we show below, this fact leads to the edge atoms tearing off upon the stress application. Formally, the energy of separation of one atom from a chain can be written as follows:

\begin{equation}
 E^b(N)=E(N)-E(N-1)-E_{at},
\end{equation}

where $E(N)$ and $E(N-1)$ are the total energies of chains with $N$ and $N-1$ atoms respectively. The regularities of change in this value with increasing of the number of atoms in carbyne are presented in Fig.~\ref{fig:fig5}. By substituting $E(N)$ and $E(N-1)$ from (7) to (8), while taking into account that  $E_2^b(N)=const$, one can see that the non-monotonic change in $E^b(N)$ value is mainly due to the change in the binding energy of the edge atom,~$E_1^b(N)$:

\begin{equation}
 E^b(N)=2\left[E_1^b(N)-E_1^b(N-1)\right]+E_{cum}^b,%
\end{equation}

At $N\geq16$ the $E_1^b(N)$ value is almost equal to $E_1^b(N-1)$. Therefore, with increasing of $N$, the value of $E^b(N)$ approaches the value of binding energy of the carbon atom in cumulene, $E_{cum}=-7.71 eV$, which agrees well with the results of \emph{ab-initio} calculations (Fig. ~\ref{fig:fig5}). Therefore, the equation (8) describes the energy balance in the separation of one atom from the chain, so it can be used to determine the binding energy in the infinite chain. However, it does not take into account specific features of interatomic interaction in the chains of finite length.

Equation (7) for the total energy of chain allows us to write down analytical expression for the average binding energy $E_0^b(N)$ per atom:

\begin{equation}
 E_0^b(N)=\frac{2\left[E_1^b(N)+E_2^b(N)-2E_{cum}^b\right]}{N}+E_{cum}^b,%
\end{equation}

When $N\geq16$ the values $E_1^b(N)$ and $E_2^b(N)$ do not depend on the number of atoms in a chain, and are equal to $E_1^b(N)=-5.80eV$ and $E_2^b(N)=-6.58eV$. In this case the average value of binding energy per atom can be found using the following expression:

\begin{equation}
 E_0^b(N)=\frac{A}{N}+E_{cum}^b,%
\end{equation}

where $A=6.08eV$.

According to Fig.~\ref{fig:fig5} this relation agrees well with the results of our ab-initio calculations and can be used to predict the average value of the binding energy per atom in carbynes containing different number of atoms. The difference between $E(N)$ for the chain of finite length and infinite chain with a cumulative-structure is due to the presence of the edges atoms (edge effect), and in case of $N=\infty$ we have $E_0^b(N)=E_{cum}^b$.

To calculate the mechanical properties of carbyne, we performed modeling of tension of chains of various lengths up to their complete break. As results of these calculations, the values of elasticity coefficient $k_Y$ and the elasticity modulus $Y$, the value of the maximum force $F_c$ as well as the respective values of critical strains of the whole chain $e_c$ (Fig. ~\ref{fig:fig1}a, b) and the edge bond $\varepsilon_c$ (Fig.~\ref{fig:fig1} c, d) were determined. Besides these values, the breaking strain of the whole chain $e_f$ (Fig ~\ref{fig:fig1} a, b) and the edge bond $\varepsilon_f$(Fig.~\ref{fig:fig1} c, d) were found. The  $F_c$ value defines the level of strength of carbine, and $e_c$ characterizes its fragility.

\begin{figure}
\includegraphics{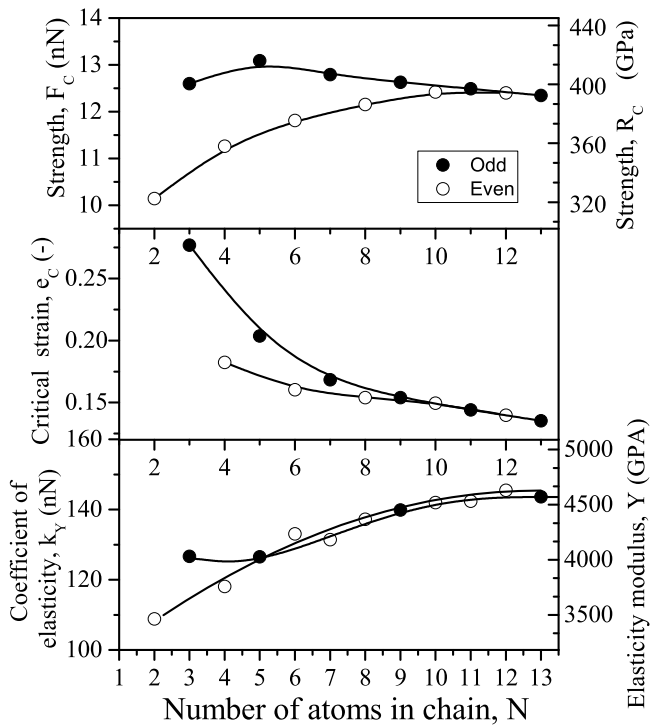}
\caption{\label{fig:fig6} The effect of the number of atoms in carbyne on its strength  $F_c$, fragility $e_c$, and hardness $k_Y$.}
\end{figure}

According to the results of our \emph{ab-initio }simulations, the magnitude of strength of the carbyne is determined by the strength of atomic bond of the edge atom. The value of this strength depends both on the total number of atoms in the chain, and on whether this number is odd or even. We observe that the strength of carbynes with an odd number of atoms is higher in comparison with the strength of even-numbered carbynes. Carbyne containing 5 atoms has maximum strength of $F_c=13.09 nN$. Chains with an odd number of atoms not only show higher strength, but are also less fragile because their instability takes place at higher values of critical strain $e_c$. The simulation results, presented in Fig. ~\ref{fig:fig6}, show that the difference between the strength of even-numbered and odd-numbered chains decreases with increase of the number of atoms in chain, and at $N\geq12$ this difference practically disappears. The increase in the number of atoms in the chains gives rise to an increase in the stiffness of the chain, $k_Y$, and to a steep (1.3 - 1.5 times) increase in their fragility (decrease in $e_c$). The latter is due to the inhomogeneous deformation of the chain, which shows itself as the localization of deformation in the bond between the first and second atoms because the edge bond is weakest. This effect is enhanced with increasing the chain length. Quantitatively, it shows itself in a significant difference between the value of fracture strain of the \emph{whole chain}, $e_f$, and the critical fracture strain of the \emph{atomic bond} between the first and second atoms $\varepsilon_f$ (Fig. ~\ref{fig:fig1}).

It should be noted that the disappearance of the "scale" and the "even-odd" effects for strength occurs at reaching of a smaller number of atoms in the chain (12 atoms) instead of 16 atoms for the interatomic interaction energy. For the coefficient of elasticity, $k_Y$, the difference between "even" and "odd" chains vanishes starting from $N\geq16$ atoms.

\begin{figure}
\includegraphics{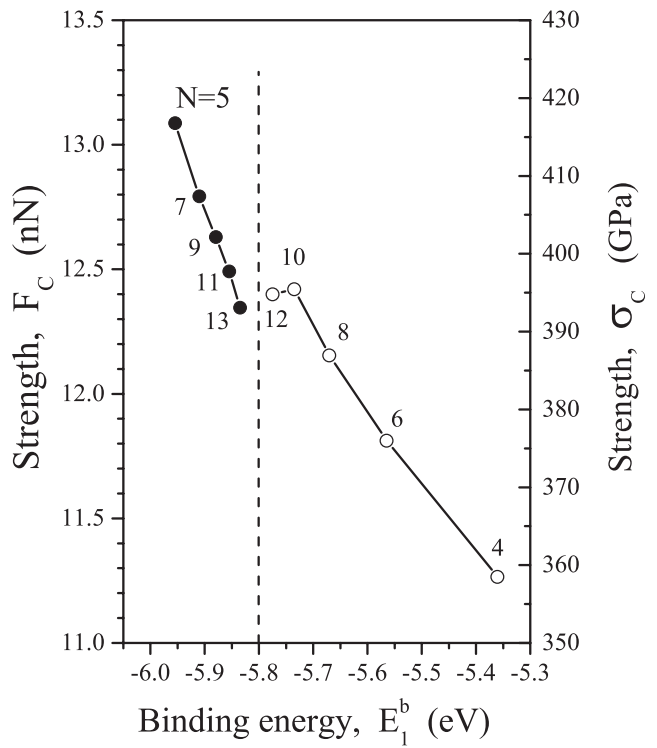}
\caption{\label{fig:fig7} The relationship between the strength of the chain, $F_c$ and the binding energy $E_1^b$ of the edge atom.}
\end{figure}

Usually the strength of interatomic bonds is estimated not directly, but using the value of binding energy. Therefore, we did a comparison between the binding energy of the edge atom $E_1^b$, and the value of the critical force of instability of the edge bond, $F_c$, which defines the strength of the whole chain. As it is shown in Fig. ~\ref{fig:fig7}, for chains containing more than 3 atoms, the increase in the binding energy of the edge atom, $E_1^b$, is accompanied by the increase in strength of the chain.

According to our results, the strength of carbyne ranges from 11.3 nN (360 GPa) to 13.1 nN (417 GPa). These values are higher than the experimental strength of 270 GPa ~\cite{Mikhailovskij18}. It is probably due to the fact that the experimental value of 270 GPa is a lower bound of strength of carbyne, as it determines the strength of the contact atomic bond between the chain and graphene from which this chain was drawn out.

Finally we should note that the different symmetry of the even-numbered and odd-numbered carbyne is the main reason for "even-odd" effect. One carbon atom is always located in the center of symmetry of the chain with an odd number of atoms. This results in the fact that the binding energy of such atom is the highest one. By virtue of the long-range interatomic interaction in the carbon chains, it causes an increase in bond strength of the edge atoms.

\section{Conclusion}
We performed a detailed theoretical study of atomic and mechanical properties of the carbyne chains using the \emph{ab initio} methods. Our results demonstrate that in carbynes with number of atoms $N\geq10$ the length of the "inner" interatomic bonds (starting from the fourth from the edge) becomes equal to the bond length in cumulenes. The existence of this "inner" structure is one of the most fundamental differences between the atomic structure of carbyne and that of the infinite chains, which show stable polyyne structure.

It is shown that unlike the carbon chains of infinite length, the interatomic distance in carbynes varies along the chain : The bond length between the first and the second (from the edge) atoms is the largest one, while the bond length between the third and the forth atoms has the smallest value. These atoms are effectively separating the inner part of carbyne from the first and second atoms. In short carbines ($N\leq16$), the distance between the first and second atoms depends both on the total number ($N$) of atoms in the carbyne, and on whether N is odd or even. With the growing number of atoms in carbyne, the length of the edge interatomic bond in even-numbered carbynes decreases, while this bond is increasing in odd-numbered carbines. When the total number of carbon atoms is $N\geq16$ , these values approach 1.306\AA, which is only 0.005\AA $ $ greater than the bond length in polyyne.

The total energy calculation showed that the energy of interatomic interaction in carbynes depends on the position of the atom in the chain. Depending on the value of the binding energy, one can distinguish three types of atoms: (i) the edge atoms that have the lowest binding energy (Å=-5.80eV for $N\geq16$); (ii) inner atoms which binding energy in chains with $N\geq10$ is equal to the binding energy in cumulenes (-7.71 eV); and (iii) the intermediate atoms with a binding energy of -6.58 eV. In short carbynes ($N\leq16$) the binding energy of the edge atoms depends on both the total number of atoms in the carbyne, and the parity of this number. The binding energy of these atoms in the even-numbered carbynes is always lower. Its value increases with the number of atoms, and in the odd-numbered carbynes, it, on the contrary, decreases.

We show that the mechanical properties of carbyne containing more than 4 and less than 12 atoms are governed by the scale and "even-odd" effects. The chains with an odd number of atoms are stronger and less fragile than the even-numbered chains. However, these effects vanish when the number of atoms in the chain is more than 10.

Carbyne is the \textbf{strongest} material in the world. Lower experimental estimation of its strength is equal to 270 GPa at 3K.  Our \emph{ab initio} calculations demonstrate that the 5-atom carbine chains show the maximum strength of 13.1 nN (417 GPa), while 392 GPa was obtained for this value for the the carbynes with $N\geq12$.

\begin{acknowledgments}
We wish to acknowledge Dr. S. Yablonovsky for the assistance in the \textit{ab initio} calculations.
\end{acknowledgments}


\bibliography{Timoshevskii}

\providecommand{\noopsort}[1]{}\providecommand{\singleletter}[1]{#1}%
\begin{thebibliography}{28}%
\makeatletter
\providecommand \@ifxundefined [1]{%
 \@ifx{#1\undefined}
}%
\providecommand \@ifnum [1]{%
 \ifnum #1\expandafter \@firstoftwo
 \else \expandafter \@secondoftwo
 \fi
}%
\providecommand \@ifx [1]{%
 \ifx #1\expandafter \@firstoftwo
 \else \expandafter \@secondoftwo
 \fi
}%
\providecommand \natexlab [1]{#1}%
\providecommand \enquote  [1]{``#1''}%
\providecommand \bibnamefont  [1]{#1}%
\providecommand \bibfnamefont [1]{#1}%
\providecommand \citenamefont [1]{#1}%
\providecommand \href@noop [0]{\@secondoftwo}%
\providecommand \href [0]{\begingroup \@sanitize@url \@href}%
\providecommand \@href[1]{\@@startlink{#1}\@@href}%
\providecommand \@@href[1]{\endgroup#1\@@endlink}%
\providecommand \@sanitize@url [0]{\catcode `\\12\catcode `\$12\catcode
  `\&12\catcode `\#12\catcode `\^12\catcode `\_12\catcode `\%12\relax}%
\providecommand \@@startlink[1]{}%
\providecommand \@@endlink[0]{}%
\providecommand \url  [0]{\begingroup\@sanitize@url \@url }%
\providecommand \@url [1]{\endgroup\@href {#1}{\urlprefix }}%
\providecommand \urlprefix  [0]{URL }%
\providecommand \Eprint [0]{\href }%
\providecommand \doibase [0]{http://dx.doi.org/}%
\providecommand \selectlanguage [0]{\@gobble}%
\providecommand \bibinfo  [0]{\@secondoftwo}%
\providecommand \bibfield  [0]{\@secondoftwo}%
\providecommand \translation [1]{[#1]}%
\providecommand \BibitemOpen [0]{}%
\providecommand \bibitemStop [0]{}%
\providecommand \bibitemNoStop [0]{.\EOS\space}%
\providecommand \EOS [0]{\spacefactor3000\relax}%
\providecommand \BibitemShut  [1]{\csname bibitem#1\endcsname}%
\let\auto@bib@innerbib\@empty
\bibitem [{\citenamefont {Lou}\ \emph {et~al.}(1995)\citenamefont {Lou},
  \citenamefont {Nordlander},\ and\ \citenamefont {Smalley}}]{Lou1}%
  \BibitemOpen
  \bibfield  {author} {\bibinfo {author} {\bibfnamefont {L.}~\bibnamefont
  {Lou}}, \bibinfo {author} {\bibfnamefont {P.}~\bibnamefont {Nordlander}}, \
  and\ \bibinfo {author} {\bibfnamefont {R.~E.}\ \bibnamefont {Smalley}},\
  }\href@noop {} {\bibfield  {journal} {\bibinfo  {journal} {Phys.\ Rev.\ B}\
  }\textbf {\bibinfo {volume} {52}},\ \bibinfo {pages} {1429} (\bibinfo {year}
  {1995})}\BibitemShut {NoStop}%
\bibitem [{\citenamefont {Ravagnan}\ \emph {et~al.}(2009)\citenamefont
  {Ravagnan}, \citenamefont {Manini}, \citenamefont {Cinquanta}, \citenamefont
  {Onida}, \citenamefont {Sangalli}, \citenamefont {Motta}, \citenamefont
  {Devetta}, \citenamefont {Bordoni}, \citenamefont {Piseri},\ and\
  \citenamefont {Milani}}]{Ravagnan2}%
  \BibitemOpen
  \bibfield  {author} {\bibinfo {author} {\bibfnamefont {L.}~\bibnamefont
  {Ravagnan}}, \bibinfo {author} {\bibfnamefont {N.}~\bibnamefont {Manini}},
  \bibinfo {author} {\bibfnamefont {E.}~\bibnamefont {Cinquanta}}, \bibinfo
  {author} {\bibfnamefont {G.}~\bibnamefont {Onida}}, \bibinfo {author}
  {\bibfnamefont {D.}~\bibnamefont {Sangalli}}, \bibinfo {author}
  {\bibfnamefont {C.}~\bibnamefont {Motta}}, \bibinfo {author} {\bibfnamefont
  {M.}~\bibnamefont {Devetta}}, \bibinfo {author} {\bibfnamefont
  {A.}~\bibnamefont {Bordoni}}, \bibinfo {author} {\bibfnamefont
  {P.}~\bibnamefont {Piseri}}, \ and\ \bibinfo {author} {\bibfnamefont
  {P.}~\bibnamefont {Milani}},\ }\href@noop {} {\bibfield  {journal} {\bibinfo
  {journal} {Phys.\ Rev.\ Lett.}\ }\textbf {\bibinfo {volume} {102}},\ \bibinfo
  {pages} {245502} (\bibinfo {year} {2009})}\BibitemShut {NoStop}%
\bibitem [{\citenamefont {Cinquanta}\ \emph {et~al.}(2011)\citenamefont
  {Cinquanta}, \citenamefont {Ravagnan}, \citenamefont {Castelli},
  \citenamefont {Cataldo}, \citenamefont {Manini}, \citenamefont {Onida},\ and\
  \citenamefont {Milani}}]{Cinquanta3}%
  \BibitemOpen
  \bibfield  {author} {\bibinfo {author} {\bibfnamefont {E.}~\bibnamefont
  {Cinquanta}}, \bibinfo {author} {\bibfnamefont {L.}~\bibnamefont {Ravagnan}},
  \bibinfo {author} {\bibfnamefont {I.~E.}\ \bibnamefont {Castelli}}, \bibinfo
  {author} {\bibfnamefont {F.}~\bibnamefont {Cataldo}}, \bibinfo {author}
  {\bibfnamefont {N.}~\bibnamefont {Manini}}, \bibinfo {author} {\bibfnamefont
  {G.}~\bibnamefont {Onida}}, \ and\ \bibinfo {author} {\bibfnamefont
  {P.}~\bibnamefont {Milani}},\ }\href@noop {} {\bibfield  {journal} {\bibinfo
  {journal} {J.\ Chem.\ Phys.}\ }\textbf {\bibinfo {volume} {135}},\ \bibinfo
  {pages} {2011} (\bibinfo {year} {2011})}\BibitemShut {NoStop}%
\bibitem [{\citenamefont {Rinzler}\ \emph
  {et~al.}(1995{\natexlab{a}})\citenamefont {Rinzler}, \citenamefont {Hafner},
  \citenamefont {Nikolaev}, \citenamefont {Lou}, \citenamefont {Kim},
  \citenamefont {Tomanek}, \citenamefont {Nordlander}, \citenamefont
  {Colbert},\ and\ \citenamefont {Smalley}}]{Rinzler4}%
  \BibitemOpen
  \bibfield  {author} {\bibinfo {author} {\bibfnamefont {G.}~\bibnamefont
  {Rinzler}}, \bibinfo {author} {\bibfnamefont {J.~H.}\ \bibnamefont {Hafner}},
  \bibinfo {author} {\bibfnamefont {P.}~\bibnamefont {Nikolaev}}, \bibinfo
  {author} {\bibfnamefont {L.}~\bibnamefont {Lou}}, \bibinfo {author}
  {\bibfnamefont {S.~G.}\ \bibnamefont {Kim}}, \bibinfo {author} {\bibfnamefont
  {D.}~\bibnamefont {Tomanek}}, \bibinfo {author} {\bibfnamefont
  {P.}~\bibnamefont {Nordlander}}, \bibinfo {author} {\bibfnamefont {D.~T.}\
  \bibnamefont {Colbert}}, \ and\ \bibinfo {author} {\bibfnamefont {R.~E.}\
  \bibnamefont {Smalley}},\ }\href@noop {} {\bibfield  {journal} {\bibinfo
  {journal} {Science}\ }\textbf {\bibinfo {volume} {269}},\ \bibinfo {pages}
  {1550} (\bibinfo {year} {1995}{\natexlab{a}})}\BibitemShut {NoStop}%
\bibitem [{\citenamefont {Durgun}\ \emph {et~al.}(2006)\citenamefont {Durgun},
  \citenamefont {Senger}, \citenamefont {Mehrez}, \citenamefont {Dag}, ,\ and\
  \citenamefont {Ciraci}}]{Durgun5}%
  \BibitemOpen
  \bibfield  {author} {\bibinfo {author} {\bibfnamefont {E.~E.}\ \bibnamefont
  {Durgun}}, \bibinfo {author} {\bibfnamefont {R.~T.}\ \bibnamefont {Senger}},
  \bibinfo {author} {\bibfnamefont {H.}~\bibnamefont {Mehrez}}, \bibinfo
  {author} {\bibfnamefont {S.}~\bibnamefont {Dag}}, , \ and\ \bibinfo {author}
  {\bibfnamefont {S.}~\bibnamefont {Ciraci}},\ }\href@noop {} {\bibfield
  {journal} {\bibinfo  {journal} {Europhys.\ Lett.}\ }\textbf {\bibinfo
  {volume} {73}},\ \bibinfo {pages} {642} (\bibinfo {year} {2006})}\BibitemShut
  {NoStop}%
\bibitem [{\citenamefont {Wang}\ \emph {et~al.}(2007)\citenamefont {Wang},
  \citenamefont {Ning}, \citenamefont {Lin}, \citenamefont {Li}, ,\ and\
  \citenamefont {Zhuang}}]{Wang6}%
  \BibitemOpen
  \bibfield  {author} {\bibinfo {author} {\bibfnamefont {Y.}~\bibnamefont
  {Wang}}, \bibinfo {author} {\bibfnamefont {X.-J.}\ \bibnamefont {Ning}},
  \bibinfo {author} {\bibfnamefont {Z.-Z.}\ \bibnamefont {Lin}}, \bibinfo
  {author} {\bibfnamefont {P.}~\bibnamefont {Li}}, , \ and\ \bibinfo {author}
  {\bibfnamefont {J.}~\bibnamefont {Zhuang}},\ }\href@noop {} {\bibfield
  {journal} {\bibinfo  {journal} {Phys.\ Rev.\ B}\ }\textbf {\bibinfo {volume}
  {76}},\ \bibinfo {pages} {165423} (\bibinfo {year} {2007})}\BibitemShut
  {NoStop}%
\bibitem [{\citenamefont {Erdogan}\ \emph {et~al.}(2011)\citenamefont
  {Erdogan}, \citenamefont {Popov}, \citenamefont {Rocha}, \citenamefont
  {Cuniberti}, \citenamefont {Roche},\ and\ \citenamefont
  {Seifert}}]{Erdogan7}%
  \BibitemOpen
  \bibfield  {author} {\bibinfo {author} {\bibfnamefont {E.}~\bibnamefont
  {Erdogan}}, \bibinfo {author} {\bibfnamefont {I.}~\bibnamefont {Popov}},
  \bibinfo {author} {\bibfnamefont {C.~G.}\ \bibnamefont {Rocha}}, \bibinfo
  {author} {\bibfnamefont {G.}~\bibnamefont {Cuniberti}}, \bibinfo {author}
  {\bibfnamefont {S.}~\bibnamefont {Roche}}, \ and\ \bibinfo {author}
  {\bibfnamefont {G.}~\bibnamefont {Seifert}},\ }\href@noop {} {\bibfield
  {journal} {\bibinfo  {journal} {Phys.\ Rev.}\ }\textbf {\bibinfo {volume}
  {83}},\ \bibinfo {pages} {041401(R)} (\bibinfo {year} {2011})}\BibitemShut
  {NoStop}%
\bibitem [{\citenamefont {Rinzler}\ \emph
  {et~al.}(1995{\natexlab{b}})\citenamefont {Rinzler}, \citenamefont {Hafner},
  \citenamefont {Nikolaev}, \citenamefont {Nordlander}, \citenamefont
  {Colbert}, \citenamefont {Smalley}, \citenamefont {Lou}, \citenamefont
  {Kim},\ and\ \citenamefont {Tomanek}}]{Rinzler8}%
  \BibitemOpen
  \bibfield  {author} {\bibinfo {author} {\bibfnamefont {G.}~\bibnamefont
  {Rinzler}}, \bibinfo {author} {\bibfnamefont {J.}~\bibnamefont {Hafner}},
  \bibinfo {author} {\bibfnamefont {P.}~\bibnamefont {Nikolaev}}, \bibinfo
  {author} {\bibfnamefont {P.}~\bibnamefont {Nordlander}}, \bibinfo {author}
  {\bibfnamefont {D.~T.}\ \bibnamefont {Colbert}}, \bibinfo {author}
  {\bibfnamefont {R.~E.}\ \bibnamefont {Smalley}}, \bibinfo {author}
  {\bibfnamefont {L.}~\bibnamefont {Lou}}, \bibinfo {author} {\bibfnamefont
  {S.~G.}\ \bibnamefont {Kim}}, \ and\ \bibinfo {author} {\bibfnamefont
  {D.}~\bibnamefont {Tomanek}},\ }\href@noop {} {\bibfield  {journal} {\bibinfo
   {journal} {Science}\ }\textbf {\bibinfo {volume} {269}},\ \bibinfo {pages}
  {1550} (\bibinfo {year} {1995}{\natexlab{b}})}\BibitemShut {NoStop}%
\bibitem [{\citenamefont {Lang}\ and\ \citenamefont {Avouris}(1998)}]{Lang9}%
  \BibitemOpen
  \bibfield  {author} {\bibinfo {author} {\bibfnamefont {N.~D.}\ \bibnamefont
  {Lang}}\ and\ \bibinfo {author} {\bibfnamefont {P.}~\bibnamefont {Avouris}},\
  }\href@noop {} {\bibfield  {journal} {\bibinfo  {journal} {Phys.\ Rev.\
  Lett.}\ }\textbf {\bibinfo {volume} {81}},\ \bibinfo {pages} {3515} (\bibinfo
  {year} {1998})}\BibitemShut {NoStop}%
\bibitem [{\citenamefont {Yazdani}(1996)}]{Yazdani10}%
  \BibitemOpen
  \bibfield  {author} {\bibinfo {author} {\bibfnamefont {N.~L.}\ \bibnamefont
  {Yazdani}, \bibfnamefont {D.~M.~Eigler}},\ }\href@noop {} {\bibfield
  {journal} {\bibinfo  {journal} {Science}\ }\textbf {\bibinfo {volume}
  {272}},\ \bibinfo {pages} {1921} (\bibinfo {year} {1996})}\BibitemShut
  {NoStop}%
\bibitem [{\citenamefont {Ragab}\ and\ \citenamefont
  {Basaran}(2011)}]{Ragab11}%
  \BibitemOpen
  \bibfield  {author} {\bibinfo {author} {\bibfnamefont {T.}~\bibnamefont
  {Ragab}}\ and\ \bibinfo {author} {\bibfnamefont {C.}~\bibnamefont
  {Basaran}},\ }\href@noop {} {\bibfield  {journal} {\bibinfo  {journal} {J.\
  Electron.\ Packag.}\ }\textbf {\bibinfo {volume} {133(2)}},\ \bibinfo {pages}
  {020903} (\bibinfo {year} {2011})}\BibitemShut {NoStop}%
\bibitem [{\citenamefont {Ataca}\ and\ \citenamefont {Ciraci}(2011)}]{Ataca12}%
  \BibitemOpen
  \bibfield  {author} {\bibinfo {author} {\bibfnamefont {C.}~\bibnamefont
  {Ataca}}\ and\ \bibinfo {author} {\bibfnamefont {S.}~\bibnamefont {Ciraci}},\
  }\href@noop {} {\bibfield  {journal} {\bibinfo  {journal} {Phys.\ Rev.\ B}\
  }\textbf {\bibinfo {volume} {83}},\ \bibinfo {pages} {235417} (\bibinfo
  {year} {2011})}\BibitemShut {NoStop}%
\bibitem [{\citenamefont {Kavan}\ \emph {et~al.}(1995)\citenamefont {Kavan},
  \citenamefont {Hlavat\'y}, \citenamefont {Kastner},\ and\ \citenamefont
  {Kuzmany}}]{Kavan13}%
  \BibitemOpen
  \bibfield  {author} {\bibinfo {author} {\bibfnamefont {L.}~\bibnamefont
  {Kavan}}, \bibinfo {author} {\bibfnamefont {J.}~\bibnamefont {Hlavat\'y}},
  \bibinfo {author} {\bibfnamefont {J.}~\bibnamefont {Kastner}}, \ and\
  \bibinfo {author} {\bibfnamefont {H.}~\bibnamefont {Kuzmany}},\ }\href@noop
  {} {\bibfield  {journal} {\bibinfo  {journal} {Carbon}\ }\textbf {\bibinfo
  {volume} {33}},\ \bibinfo {pages} {1321} (\bibinfo {year}
  {1995})}\BibitemShut {NoStop}%
\bibitem [{\citenamefont {Casari}\ \emph {et~al.}(2004)\citenamefont {Casari},
  \citenamefont {Bassi}, \citenamefont {Ravagnan}, \citenamefont {Siviero},
  \citenamefont {Lenardi}, \citenamefont {Piseri}, \citenamefont {Bongiorno},
  \citenamefont {Bottani},\ and\ \citenamefont {Milani}}]{Casari14}%
  \BibitemOpen
  \bibfield  {author} {\bibinfo {author} {\bibfnamefont {C.~S.}\ \bibnamefont
  {Casari}}, \bibinfo {author} {\bibfnamefont {A.~L.}\ \bibnamefont {Bassi}},
  \bibinfo {author} {\bibfnamefont {L.}~\bibnamefont {Ravagnan}}, \bibinfo
  {author} {\bibfnamefont {F.}~\bibnamefont {Siviero}}, \bibinfo {author}
  {\bibfnamefont {C.}~\bibnamefont {Lenardi}}, \bibinfo {author} {\bibfnamefont
  {P.}~\bibnamefont {Piseri}}, \bibinfo {author} {\bibfnamefont
  {G.}~\bibnamefont {Bongiorno}}, \bibinfo {author} {\bibfnamefont {C.~E.}\
  \bibnamefont {Bottani}}, \ and\ \bibinfo {author} {\bibfnamefont
  {P.}~\bibnamefont {Milani}},\ }\href@noop {} {\bibfield  {journal} {\bibinfo
  {journal} {Phys.\ Rev.\ B}\ }\textbf {\bibinfo {volume} {69}},\ \bibinfo
  {pages} {075422} (\bibinfo {year} {2004})}\BibitemShut {NoStop}%
\bibitem [{\citenamefont {Jin}\ \emph {et~al.}(2009)\citenamefont {Jin},
  \citenamefont {Lan}, \citenamefont {Peng}, \citenamefont {Suenaga},\ and\
  \citenamefont {Iijima}}]{Jin15}%
  \BibitemOpen
  \bibfield  {author} {\bibinfo {author} {\bibfnamefont {C.}~\bibnamefont
  {Jin}}, \bibinfo {author} {\bibfnamefont {H.}~\bibnamefont {Lan}}, \bibinfo
  {author} {\bibfnamefont {L.}~\bibnamefont {Peng}}, \bibinfo {author}
  {\bibfnamefont {K.}~\bibnamefont {Suenaga}}, \ and\ \bibinfo {author}
  {\bibfnamefont {S.}~\bibnamefont {Iijima}},\ }\href@noop {} {\bibfield
  {journal} {\bibinfo  {journal} {Phys.\ Rev.\ Lett.}\ }\textbf {\bibinfo
  {volume} {102}},\ \bibinfo {pages} {205501} (\bibinfo {year}
  {2009})}\BibitemShut {NoStop}%
\bibitem [{\citenamefont {Liu}\ \emph {et~al.}(2013)\citenamefont {Liu},
  \citenamefont {Artyukhov}, \citenamefont {Lee}, \citenamefont {Xu}, ,\ and\
  \citenamefont {Yakobson}}]{Liu16}%
  \BibitemOpen
  \bibfield  {author} {\bibinfo {author} {\bibfnamefont {M.}~\bibnamefont
  {Liu}}, \bibinfo {author} {\bibfnamefont {V.~I.}\ \bibnamefont {Artyukhov}},
  \bibinfo {author} {\bibfnamefont {H.}~\bibnamefont {Lee}}, \bibinfo {author}
  {\bibfnamefont {F.}~\bibnamefont {Xu}}, , \ and\ \bibinfo {author}
  {\bibfnamefont {B.~I.}\ \bibnamefont {Yakobson}},\ }\href@noop {} {\bibfield
  {journal} {\bibinfo  {journal} {ACS Nano}\ }\textbf {\bibinfo {volume} {7
  (11)}},\ \bibinfo {pages} {10075} (\bibinfo {year} {2013})}\BibitemShut
  {NoStop}%
\bibitem [{\citenamefont {Kotrechko}\ \emph {et~al.}(2012)\citenamefont
  {Kotrechko}, \citenamefont {Mazilov}, \citenamefont {Mazilova}, \citenamefont
  {Sadanov},\ and\ \citenamefont {Mikhailovskij}}]{Kotrechko17}%
  \BibitemOpen
  \bibfield  {author} {\bibinfo {author} {\bibfnamefont {S.}~\bibnamefont
  {Kotrechko}}, \bibinfo {author} {\bibfnamefont {A.~A.}\ \bibnamefont
  {Mazilov}}, \bibinfo {author} {\bibfnamefont {T.~I.}\ \bibnamefont
  {Mazilova}}, \bibinfo {author} {\bibfnamefont {E.~V.}\ \bibnamefont
  {Sadanov}}, \ and\ \bibinfo {author} {\bibfnamefont {I.~M.}\ \bibnamefont
  {Mikhailovskij}},\ }\href@noop {} {\bibfield  {journal} {\bibinfo  {journal}
  {Tech.\ Phys.\ Lett.}\ }\textbf {\bibinfo {volume} {38}},\ \bibinfo {pages}
  {132} (\bibinfo {year} {2012})}\BibitemShut {NoStop}%
\bibitem [{\citenamefont {Mikhailovskij}\ \emph {et~al.}(2013)\citenamefont
  {Mikhailovskij}, \citenamefont {Sadanov}, \citenamefont {Kotrechko},
  \citenamefont {Ksenofontov},\ and\ \citenamefont
  {Mazilova}}]{Mikhailovskij18}%
  \BibitemOpen
  \bibfield  {author} {\bibinfo {author} {\bibfnamefont {I.}~\bibnamefont
  {Mikhailovskij}}, \bibinfo {author} {\bibfnamefont {E.}~\bibnamefont
  {Sadanov}}, \bibinfo {author} {\bibfnamefont {S.}~\bibnamefont {Kotrechko}},
  \bibinfo {author} {\bibfnamefont {V.}~\bibnamefont {Ksenofontov}}, \ and\
  \bibinfo {author} {\bibfnamefont {T.}~\bibnamefont {Mazilova}},\ }\href@noop
  {} {\bibfield  {journal} {\bibinfo  {journal} {Phys.\ Rev.\ B}\ }\textbf
  {\bibinfo {volume} {87}},\ \bibinfo {pages} {045410} (\bibinfo {year}
  {2013})}\BibitemShut {NoStop}%
\bibitem [{\citenamefont {Huang}\ \emph {et~al.}(2006)\citenamefont {Huang},
  \citenamefont {Wu},\ and\ \citenamefont {Hwang}}]{Huang19}%
  \BibitemOpen
  \bibfield  {author} {\bibinfo {author} {\bibfnamefont {Y.}~\bibnamefont
  {Huang}}, \bibinfo {author} {\bibfnamefont {J.}~\bibnamefont {Wu}}, \ and\
  \bibinfo {author} {\bibfnamefont {K.~C.}\ \bibnamefont {Hwang}},\ }\href@noop
  {} {\bibfield  {journal} {\bibinfo  {journal} {Phys.\ Rev.\ B}\ }\textbf
  {\bibinfo {volume} {74}},\ \bibinfo {pages} {245413} (\bibinfo {year}
  {2006})}\BibitemShut {NoStop}%
\bibitem [{\citenamefont {Fan}\ \emph {et~al.}(2009)\citenamefont {Fan},
  \citenamefont {Liu}, \citenamefont {Lin}, \citenamefont {Shen},\ and\
  \citenamefont {Kuo}}]{Fan20}%
  \BibitemOpen
  \bibfield  {author} {\bibinfo {author} {\bibfnamefont {X.~F.}\ \bibnamefont
  {Fan}}, \bibinfo {author} {\bibfnamefont {L.}~\bibnamefont {Liu}}, \bibinfo
  {author} {\bibfnamefont {J.}~\bibnamefont {Lin}}, \bibinfo {author}
  {\bibfnamefont {Z.~X.}\ \bibnamefont {Shen}}, \ and\ \bibinfo {author}
  {\bibfnamefont {J.-L.}\ \bibnamefont {Kuo}},\ }\href@noop {} {\bibfield
  {journal} {\bibinfo  {journal} {ACS Nano}\ }\textbf {\bibinfo {volume}
  {11}},\ \bibinfo {pages} {3788} (\bibinfo {year} {2009})}\BibitemShut
  {NoStop}%
\bibitem [{\citenamefont {Cahangirov}\ \emph {et~al.}(2010)\citenamefont
  {Cahangirov}, \citenamefont {Topsakal},\ and\ \citenamefont
  {Ciraci}}]{Cahangirov21}%
  \BibitemOpen
  \bibfield  {author} {\bibinfo {author} {\bibfnamefont {S.}~\bibnamefont
  {Cahangirov}}, \bibinfo {author} {\bibfnamefont {M.}~\bibnamefont
  {Topsakal}}, \ and\ \bibinfo {author} {\bibfnamefont {S.}~\bibnamefont
  {Ciraci}},\ }\href@noop {} {\bibfield  {journal} {\bibinfo  {journal} {Phys.\
  Rev.\ B}\ }\textbf {\bibinfo {volume} {82}},\ \bibinfo {pages} {195444}
  (\bibinfo {year} {2010})}\BibitemShut {NoStop}%
\bibitem [{\citenamefont {Giannozzi}\ \emph {et~al.}(2009)\citenamefont
  {Giannozzi}, \citenamefont {Baroni}, \citenamefont {Bonini}, \citenamefont
  {Calandra}, \citenamefont {R.~Car}, \citenamefont {Ceresoli}, \citenamefont
  {Chiarotti}, \citenamefont {Cococcioni}, \citenamefont {Dabo},\ and\
  \citenamefont {et~al.}}]{Giannozzi22}%
  \BibitemOpen
  \bibfield  {author} {\bibinfo {author} {\bibfnamefont {P.}~\bibnamefont
  {Giannozzi}}, \bibinfo {author} {\bibfnamefont {S.}~\bibnamefont {Baroni}},
  \bibinfo {author} {\bibfnamefont {N.}~\bibnamefont {Bonini}}, \bibinfo
  {author} {\bibfnamefont {M.}~\bibnamefont {Calandra}}, \bibinfo {author}
  {\bibfnamefont {C.~C.}\ \bibnamefont {R.~Car}}, \bibinfo {author}
  {\bibfnamefont {D.}~\bibnamefont {Ceresoli}}, \bibinfo {author}
  {\bibfnamefont {G.~L.}\ \bibnamefont {Chiarotti}}, \bibinfo {author}
  {\bibfnamefont {M.}~\bibnamefont {Cococcioni}}, \bibinfo {author}
  {\bibfnamefont {I.}~\bibnamefont {Dabo}}, \ and\ \bibinfo {author}
  {\bibnamefont {et~al.}},\ }\href@noop {} {\bibfield  {journal} {\bibinfo
  {journal} {J.\ Phys.:\ Condens.\ Matter}\ }\textbf {\bibinfo {volume} {21}},\
  \bibinfo {pages} {395502} (\bibinfo {year} {2009})}\BibitemShut {NoStop}%
\bibitem [{\citenamefont {Giannozzi}\ \emph {et~al.}()\citenamefont
  {Giannozzi}, \citenamefont {Alfe}, \citenamefont {Alkauskas},\ and\
  \citenamefont {et~al.}}]{Giannozzi23}%
  \BibitemOpen
  \bibfield  {author} {\bibinfo {author} {\bibfnamefont {P.}~\bibnamefont
  {Giannozzi}}, \bibinfo {author} {\bibfnamefont {D.}~\bibnamefont {Alfe}},
  \bibinfo {author} {\bibfnamefont {A.}~\bibnamefont {Alkauskas}}, \ and\
  \bibinfo {author} {\bibnamefont {et~al.}},\ }\href@noop {} {\bibinfo
  {journal} {http://www.quantum-espresso.org}\ }\BibitemShut {NoStop}%
\bibitem [{\citenamefont {Blaha}\ \emph {et~al.}(2001)\citenamefont {Blaha},
  \citenamefont {Schwarz}, \citenamefont {Madsen}, \citenamefont {Kvasnicka},\
  and\ \citenamefont {Luitz}}]{Blaha24}%
  \BibitemOpen
\bibfield  {journal} {  }\bibfield  {author} {\bibinfo {author} {\bibfnamefont
  {P.}~\bibnamefont {Blaha}}, \bibinfo {author} {\bibfnamefont
  {K.}~\bibnamefont {Schwarz}}, \bibinfo {author} {\bibfnamefont
  {G.}~\bibnamefont {Madsen}}, \bibinfo {author} {\bibfnamefont
  {D.}~\bibnamefont {Kvasnicka}}, \ and\ \bibinfo {author} {\bibfnamefont
  {J.}~\bibnamefont {Luitz}},\ }\href@noop {} {\emph {\bibinfo {title} {WIEN2K,
  An Augmented Plane Wave + Local Orbital's Program for Calculating Crystal
  Properties}}}\ (\bibinfo  {publisher} {Technische Universitat Wien},\
  \bibinfo {address} {Austria},\ \bibinfo {year} {2001})\ \bibinfo {note} {iSBN
  3-9501031-1-2}\BibitemShut {NoStop}%
\bibitem [{\citenamefont {Perdew}\ \emph {et~al.}(1996)\citenamefont {Perdew},
  \citenamefont {Burke},\ and\ \citenamefont {Ernzerhof}}]{Perdew25}%
  \BibitemOpen
  \bibfield  {author} {\bibinfo {author} {\bibnamefont {Perdew}}, \bibinfo
  {author} {\bibfnamefont {K.}~\bibnamefont {Burke}}, \ and\ \bibinfo {author}
  {\bibfnamefont {M.}~\bibnamefont {Ernzerhof}},\ }\href@noop {} {\bibfield
  {journal} {\bibinfo  {journal} {Phys.\ Rev.\ Lett.}\ }\textbf {\bibinfo
  {volume} {77}},\ \bibinfo {pages} {3865} (\bibinfo {year}
  {1996})}\BibitemShut {NoStop}%
\bibitem [{\citenamefont {Mazilova}\ \emph {et~al.}(2010)\citenamefont
  {Mazilova}, \citenamefont {Sadanov}, \citenamefont {Ksenofontov},\ and\
  \citenamefont {Mikhailovskij}}]{Mazilova26}%
  \BibitemOpen
  \bibfield  {author} {\bibinfo {author} {\bibfnamefont {T.~I.}\ \bibnamefont
  {Mazilova}}, \bibinfo {author} {\bibfnamefont {E.~V.}\ \bibnamefont
  {Sadanov}}, \bibinfo {author} {\bibfnamefont {V.~A.}\ \bibnamefont
  {Ksenofontov}}, \ and\ \bibinfo {author} {\bibfnamefont {I.~M.}\ \bibnamefont
  {Mikhailovskij}},\ }\href@noop {} {\bibfield  {journal} {\bibinfo  {journal}
  {International\ Journal\ of\ Nanoscience}\ }\textbf {\bibinfo {volume}
  {9(3)}},\ \bibinfo {pages} {151} (\bibinfo {year} {2010})}\BibitemShut
  {NoStop}%
\bibitem [{\citenamefont {Wu}\ and\ \citenamefont {Dong}(2005)}]{Gang27}%
  \BibitemOpen
  \bibfield  {author} {\bibinfo {author} {\bibfnamefont {G.}~\bibnamefont
  {Wu}}\ and\ \bibinfo {author} {\bibfnamefont {J.}~\bibnamefont {Dong}},\
  }\href@noop {} {\bibfield  {journal} {\bibinfo  {journal} {Phys.\ Rev.}\
  }\textbf {\bibinfo {volume} {71}},\ \bibinfo {pages} {115410} (\bibinfo
  {year} {2005})}\BibitemShut {NoStop}%
\bibitem [{\citenamefont {Artyukhov}\ and\ \citenamefont
  {Liu}(2013)}]{Artyukhov28}%
  \BibitemOpen
  \bibfield  {author} {\bibinfo {author} {\bibfnamefont {V.~I.}\ \bibnamefont
  {Artyukhov}}\ and\ \bibinfo {author} {\bibfnamefont {M.}~\bibnamefont
  {Liu}},\ }\href@noop {} {}\bibinfo {howpublished} {e-print
  arXiv:cond-mat.mtrl-sci/1302.7250v2} (\bibinfo {year} {2013})\BibitemShut
  {NoStop}%
\end{thebibliography}%

\end{document}